\documentclass[twocolumn,superscriptaddress,nofootinbib,floatfix,aps,prb,citeautoscript,longbibliography]{revtex4-2}
\usepackage[utf8]{inputenc}
\usepackage[T1]{fontenc}
\usepackage{lineno}

\usepackage{hyperref}
\usepackage{graphicx}
\usepackage{color}
\usepackage{array}
\usepackage{dcolumn}
\usepackage{bm}
\usepackage{multirow}
\usepackage[version=4]{mhchem}
\usepackage{cleveref}
\usepackage{float}
\restylefloat{table}
\usepackage{makecell}
\usepackage{booktabs}

\usepackage{siunitx}
\DeclareSIUnit\atom{atom}
\DeclareSIUnit\ang{\angstrom}
\DeclareSIPostPower\3{3}
\newcommand{\etal}{\emph{et~al.}}

\begin{document}

\newcommand{\halle}{Institut für Physik, Martin-Luther-Universität
  Halle-Wittenberg, D-06099 Halle, Germany}
\newcommand{\jena}{Institut f\"ur Festk\"orpertheorie und -optik,
  Friedrich-Schiller-Universit\"at Jena, Max-Wien-Platz 1, 07743 Jena, Germany}
\author{Noah Hoffmann}
\affiliation{\halle} 
\author{Jonathan Schmidt}
\affiliation{\jena} 
\affiliation{\halle} 
\author{Silvana Botti}
\affiliation{\jena}
\author{Miguel A. L. Marques} 
\email{miguel.marques@physik.uni-halle.de}
\affiliation{\halle} 

\date{\today}

\title{Transfer learning on large datasets for the accurate prediction of material properties}
\begin{abstract}
Graph neural networks trained on large crystal structure databases are extremely effective in replacing \textit{ab initio} calculations in the discovery and characterization of materials. However, crystal structure datasets comprising millions of materials exist only for the Perdew-Burke-Ernzerhof (PBE) functional. In this work, we investigate the effectiveness of transfer learning to extend these models to other density functionals. We show that pre-training significantly reduces the size of the dataset required to achieve chemical accuracy and beyond.
We also analyze in detail the relationship between the transfer-learning performance and the size of the datasets used for the initial training of the model and transfer learning. We confirm a linear dependence of the error on the size of the datasets on a log-log scale, with a similar slope for both training and the pre-training datasets. This shows that further increasing the size of the pre-training dataset, i.e. performing additional calculations with a low-cost functional, is also effective, through transfer learning, in improving machine-learning predictions with the quality of a more accurate, and possibly computationally more involved functional.  Lastly, we compare the efficacy of interproperty and intraproperty transfer learning. 
\end{abstract}

\maketitle

\section{Introduction}

Over the past decade, machine learning models have emerged as unbeatable tools to accelerate materials research in fields ranging from quantum chemistry~\cite{chemistryreview} to drug discovery~\cite{drugreview} to solid-state materials science~\cite{ourreview, roadmap, materialsreview, materialsreview2}. One of the major drivers of progress has been the transition from simpler models based on a single material family to more advanced and more universal models~\cite{megnet, 68crystalgraphconvolution, chen2022universal}. The current state of the art are graph neural networks and graph transformers~\cite{fu2022forces}. However, the accuracy of these complex models, which include millions of parameters, depends heavily on the amount and quality of the training data~\cite{dunn2020benchmarking}. Consequently, the development of such models depends on the availability of large databases of calculations obtained with consistent computational parameters.

In quantum chemistry, there are a variety of large databases that gather calculations with varying degrees of accuracy, from density functional theory (with different functionals) to coupled-cluster~\cite{smith2020ani}. In comparison, in solid-state materials science, all large databases (>10$^5$\ compounds) contain calculations performed using the Perdew-Burke-Ernzerhof (PBE) functional~\cite{PBE}. The list includes AFLOW~\cite{aflowlib}, the OQMD~\cite{OQMD,OQMD2}, the Materials Project~\cite{materialsproject}, and DCGAT~\cite{CGATHT}. An exception is the smaller JARVIS database that encompasses $\sim 55$k calculations obtained using OptB88-vdW~\cite{PhysRevB.76.125112, PhysRevB.83.195131} and the modified Becke-Johnson potential~\cite{10.1038/s41524-020-00440-1,10.1038/sdata.2018.82,10.1103/PhysRevB.98.014107}.
The first large databases beyond the PBE functional have been published only recently~\cite{dataset, kingsbury2022performance}. These datasets are based on the PBE functional for solids (PBEsol) ~\cite{pbe-sol}, the highly constrained and appropriately normalized semilocal functional (SCAN) ~\cite{SCAN}, and the R2SCAN functional~\cite{R2Scan}. All of these density functionals yield much more accurate geometries than the PBE, and the latter two also yield  more accurate formation energies and band gaps. Nevertheless, these databases are one to two orders of magnitude smaller than the largest PBE databases. 

Currently, graph networks ~\cite{68crystalgraphconvolution, icgcnn, megnet, goodall2019predicting,CGAT} are the best performing models for datasets including more than 10$^4$\ compounds~\cite{dunn2020benchmarking}. However, they improve dramatically with increasing dataset size beyond this number.
Consequently, despite the higher accuracy of PBEsol or SCAN, models trained on these smaller datasets have a significantly larger prediction error compared to their PBE counterparts.
One way to circumvent the problem of data sparsity is to perform transfer learning and pre-training, as these approaches have proved to be extremely effective in other fields~\cite{tan2018survey,kalyan2021ammus}. 
In the areas of computer vision and natural language processing, for instance, almost all non-edge applications can be improved by using large pre-trained models~\cite{tan2018survey,kalyan2021ammus}. 

In recent years, transfer learning and multi-fidelity learning have also arrived in materials science~\cite{hutchinson2017overcoming,jha2019enhancing,smith2019approaching,Shufeng,Gupta2021,shotgun,concreteFord2022,opencatalysttransfer,chen2021disordered,oqmdcnnrf,MODNET,megnet,AtomSets}.
The published works deal with rather small datasets for both pre-training and transfer learning, usually $\lesssim10^4$\ data points for the transfer dataset and $\lesssim10^5$\ data points for the pre-training dataset. In this context, band gaps~\cite{hutchinson2017overcoming,megnet,chen2021disordered,MODNET} and formation energies~\cite{jha2019enhancing,MODNET} are the most popular features for transfer learning, since there is abundance of multi-fidelity theoretical and  experimental data ($\sim 10^3$ measurements).

Very few applications have been made in the realm of big data, since hardly any large data sets exist. Smith \etal\ \cite{smith2020ani} improved ANI~\cite{smith2019approaching} quantum chemistry force fields beyond the accuracy of DFT by transferring from the ANI-1x DFT dataset that contains 5.2~million molecules to a dataset of 0.5~million molecules computed with coupled cluster~\cite{smith2020ani}.
Kolluru \etal\ \cite{opencatalysttransfer} used a model pre-trained on the Open Catalyst Dataset OC20~\cite{oc20} to obtain better results on the smaller MD17 dataset~\cite{md17}. They successfully reduced training times and arrived at a mean improvement of $\sim$6\% over a model trained from scratch.

In this work, we investigate the benefits of transfer learning when large materials datasets are available, investigating both cross-property transfer and prediction improvement through high-accuracy data. 

We focus mainly on properties that are important for the discovery of new crystal structures. In this context, the energy distance to the convex hull of thermodynamic stability is a key quantity, as it compares the formation energy of a crystalline compound with the combined energy of the available decomposition channels. 
In Ref. we have published a data set with 175k SCAN~\cite{SCAN} total energies and PBEsol~\cite{pbe-sol} geometries of stable and metastable systems. In preparation of this work, we have extended that dataset to include additional $\sim$50k calculations of randomly selected unstable systems with a distance from the convex hull of thermodynamic stability below 800~meV/atom. Calculations using SCAN show about half the mean absolute error (MAE) for formation energies compared to calculations with the standard PBE functional~\cite{SCAN_S}. Similarly, calculations using PBEsol reduce the mean absolute percent errors on volumes by 40\% compared to PBE~\cite{latticejo}.

\begin{table*}[ht]
    \centering
    \begin{tabular}{r|c|ccc|ccc}
        & PBE &  \multicolumn{3}{|c|}{PBEsol}  & \multicolumn{3}{c}{SCAN}  \\
        &  & no transfer & only residual& full transfer&  no transfer & only residual& full transfer  \\ \hline
        E$_{\text{Hull}}$[meV/atom] & 23 & 26  & 22 (15\%)  & \textbf{19} (27\%) & 31 & 26 (16\%) & \textbf{22} (29\%) \\
        E$_{\text{Form}}$[meV/atom] & 18 & 20 & 18 (10\%) & \textbf{13} (35\%) & 24 & 22 (8\%) & \textbf{16} (33\%) \\
        Volume [\AA$^3$/atoms ] & 0.24 &0.21 & 0.18 (14\%)  & \textbf{0.16} (23\%) & & &   \\
        Band gap $\left[\si{\eV}\right]$ & 0.020 && &&0.078& 0.93 (-19\%)& \textbf{0.068} (13\%)
    \end{tabular}
    \caption{\label{tab:transfer-summary-1}Intra-property transfer learning. We report the mean absolute errors on the test set for the neural networks trained on the large PBE dataset only and the neural networks trained on the PBEsol and SCAN datasets with and without transfer learning. The different approaches for transfer learning (only residual and full transfer) are explained in the text. The models with the best performance for PBEsol/SCAN are indicated with bold letters. The percent improvement in comparison to the case of no transfer learning are shown in parenthesis.}
\end{table*}

In the following, we will demonstrate that pre-training using a large PBE dataset allows us to obtain improved predictions, with the accuracy of more advanced density functionals, even when the available training data is limited for the latter. Furthermore, we will evaluate the dependence of the error on the size of the training and pre-training data to quantify potential gains through future expansion of the datasets of calculations.
We also investigate to which extent transfer learning is useful to improve predictions of different materials properties.

\section{Results}

We started our transfer learning experiments by training crystal graph-attention neural networks~\cite{CGAT} on a PBE dataset with 1.8M structures~\cite{CGATHT} from the DCGAT database, and on the extended PBEsol and SCAN datasets from Ref.~\cite{dataset}. 
The DCGAT dataset combines compatible data from AFLOW~\cite{aflowlib}, the materials project~\cite{materialsproject} and Refs.~\cite{jonathan2018, schmidt2017, CGAT,CGATHT,garnet}.
We generally used a train/val/test split of 80/10/10\%. The validation and test set were randomly selected once and then kept constant for all experiments on SCAN and PBEsol to ensure a fair comparison. 
As a starting point, one model was trained for each of the three functionals and for four properties, specifically the distance to the convex hull (E$_{\text{Hull}}$), the formation energy (E$_{\text{Form}}$), the band gap and the volume per atom of the unit cell. For the band gaps only PBE and SCAN models were trained, as PBEsol and PBE band gaps are very similar, and we did not perform transfer learning for SCAN volumes as we only have calculated PBE and PBEsol geometries.

\begin{figure*}[htb]
  \begin{tabular}{c c}
  \centering
  \includegraphics[width=0.95\columnwidth]{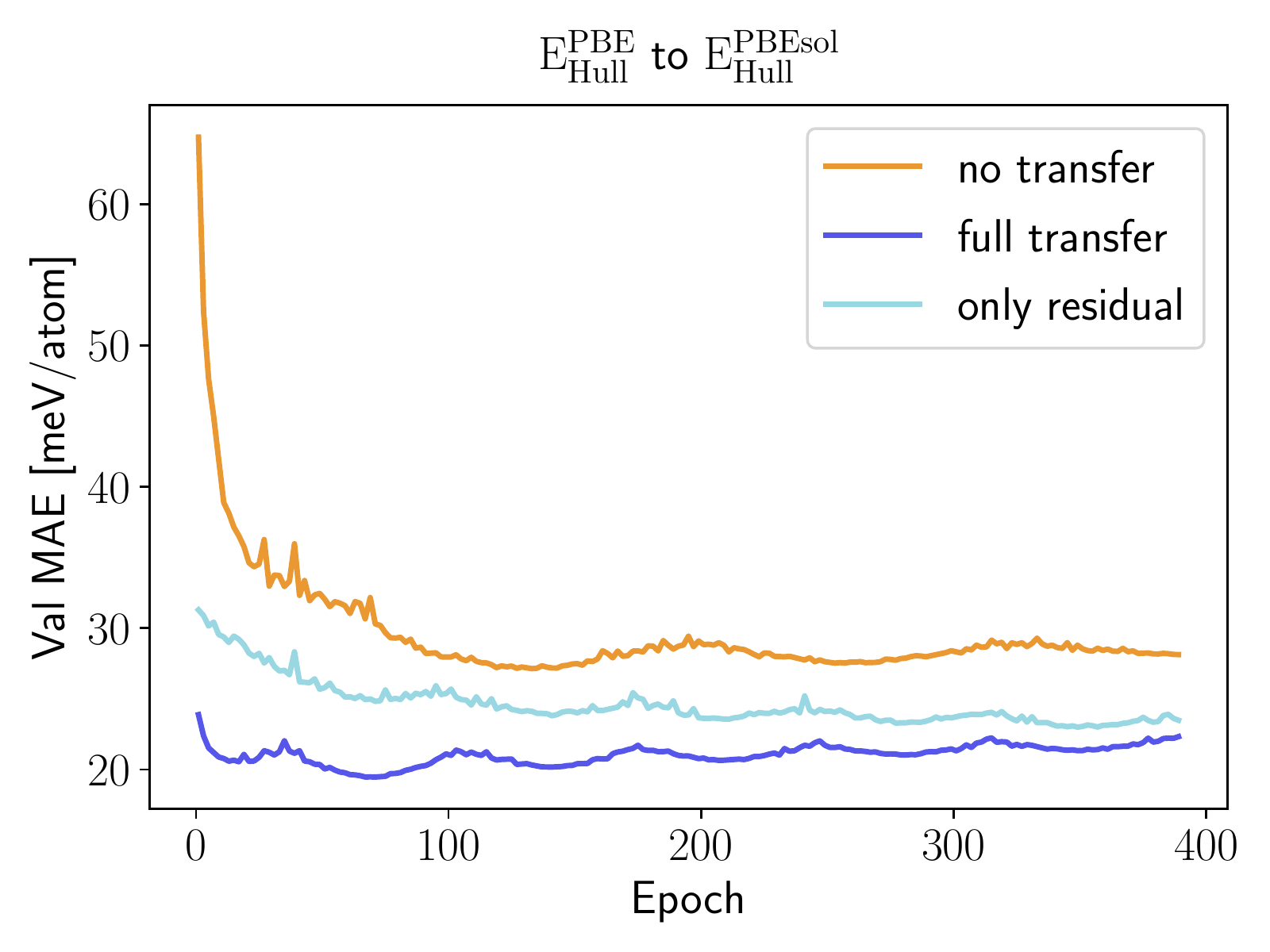}
  & \includegraphics[width=0.95\columnwidth]{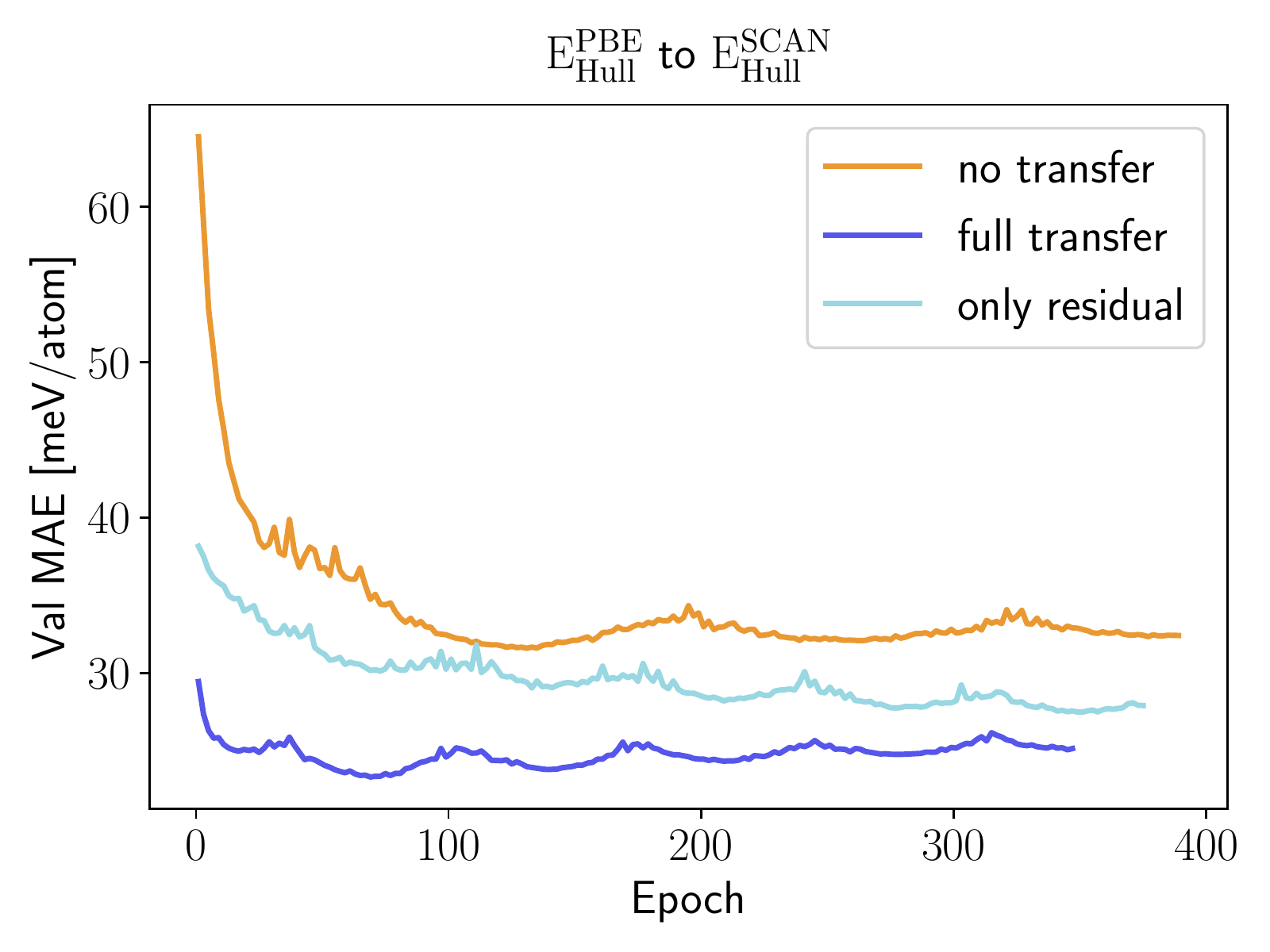}
  \end{tabular}
  \caption{Learning curves, i.e. MAE on the validation set as a function of the training epoch, for the prediction of E$_\text{Hull}^{\text{PBEsol}}$ (left) and E$_\text{Hull}^{\text{SCAN}}$ (right) using the different training procedures, without pre-training (no transfer), with pre-training using E$_\text{Hull}^{\text{PBE}}$ data and freezing part of the weights (only residual) and with pre-training using E$_\text{Hull}^{\text{PBE}}$ data and successive full reoptimization of all weights (full transfer).}
    \label{fig:e-hull-transfer}
\end{figure*}

The initial neural network, trained on the PBE datasets comprising 1.8M calculations, predicts the distance to the convex hull of the systems in the test set with a MAE of 23~meV/atom. As expected, the same neural networks, trained from scratch using significantly smaller PBEsol and SCAN datasets (this procedure is labelled ``no transfer'' in the figures and tables), perform worse and yield, respectively, 9\% and  23\% higher MAEs. To take advantage of transfer learning we considered two options: starting from the neural network trained with the large PBE dataset, we either continue the training of the whole network with the PBEsol or the SCAN data (this procedure is denoted ``full transfer''), or we fix most of the weights and train only the residual part of the neural network, i.e. the one that calculates the scalar output from the final graph embedding (this procedure is indicated as ``only residual'').
Comparing in \cref{fig:e-hull-transfer} and \cref{tab:transfer-summary-1} these two approaches of transfer learning to the original model trained solely on the smaller datasets, we can immediately conclude that, in all cases, transfer learning enhances considerably the performance of the neural network. 

Applying the neural network trained on the PBE data, after retraining only its residual part on the PBEsol/SCAN dataset, resulted in improved predictions in comparison to the ones of the same network trained only on the PBEsol/SCAN dataset. This type of transfer learning leads to a \SI{15}{\percent} smaller test MAE for the PBEsol data and \SI{16}{\percent} smaller test MAE for the SCAN data.
Retraining the full network resulted in an even better performance, with a reduction of the MAE on the test set, after only $\sim$70 epochs, by \SI{27}{\percent} and \SI{29}{\percent} for the PBEsol and SCAN data, respectively. 


If we use the same neural network to predict the formation energy instead of the distance to the convex hull, the results are very similar, as shown in \cref{tab:transfer-summary-1}. Transfer learning with retraining of the residual network leads in this case to an error reduction of \SI{10}{\percent} and \SI{8}{\percent} for PBEsol and SCAN, respectively. On the other hand, extending the training of the full network to the additional datasets brings an even higher error reduction of \SI{35} and \SI{33}{\percent} for PBEsol and SCAN, respectively. Again both transfer learning approaches give consistently a lower MAE than the neural network trained anew on smaller datasets, demonstrating the benefit of exploiting the larger database of less accurate, but computationally more affordable, PBE calculations.

The visible discrepancies in the performance of the fully retrained network and the partially retrained  network, where only the weights of the residual network are further optimized, hint to the fact that the crystal graph embeddings for the three functionals must differ significantly.



We only have available PBEsol volumes, as the data was generated with the approach of Ref.~\cite{dataset}.
Consequently, we only test the transfer learning for the PBEsol functional achieving an improvement of \SI{23}{\percent}.

The band gaps are the sole material property where we only obtain an improvement (with a 13\% MAE reduction) when the training on the new data is performed for the whole network. We have to be careful in considering MAE for this dataset as most of the materials are metals with a band gap equal to zero. If the machine predicts a metal the associated error on the band gap value will be zero, while a finite error can be associated to the prediction of an open band gap. There are remarkable differences in the band gap distribution for calculations performed with the two different functionals (also visible in \cref{fig:hist}d), as PBE underestimates the band gap more strongly than SCAN. This results in an average band gap of 0.08\,eV in the PBE dataset and of 0.47\,eV in the SCAN dataset. The fact that PBE misclassifies some semiconductors as metals leads to a MAE artificially smaller for the machine trained on PBE data, as more metals are contained in the PBE dataset. While the distributions for other properties also differ, there is in this case a  qualitative difference between metals and semiconductors and it is necessary that the network trained on the SCAN data also learns also how to distinguish false metals in the PBE dataset. The training of the residual network only is therefore totally insufficient. 

\begin{figure*}
    \centering
    \includegraphics[width=.45\linewidth]{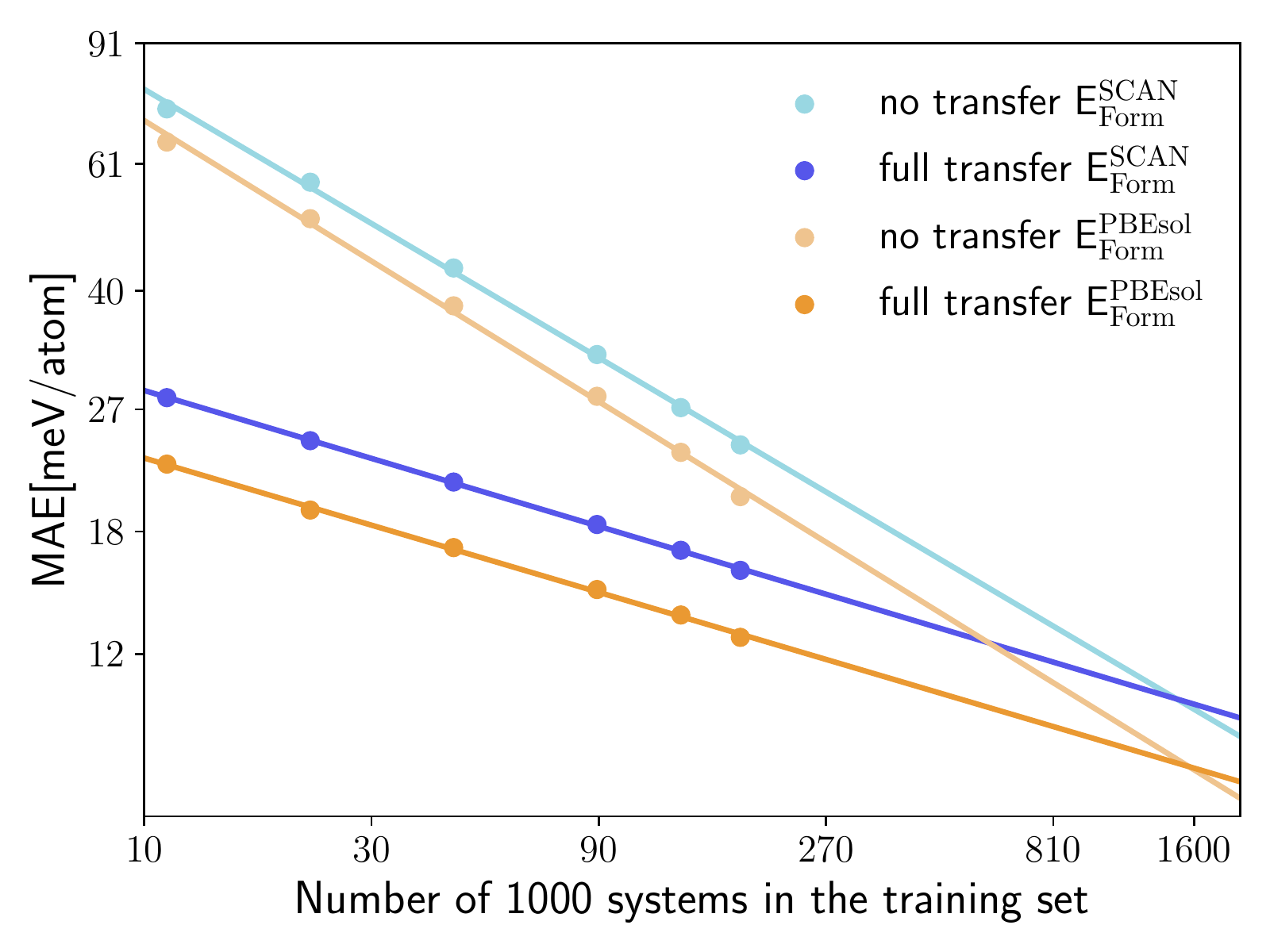}
    \includegraphics[width=.45\linewidth]{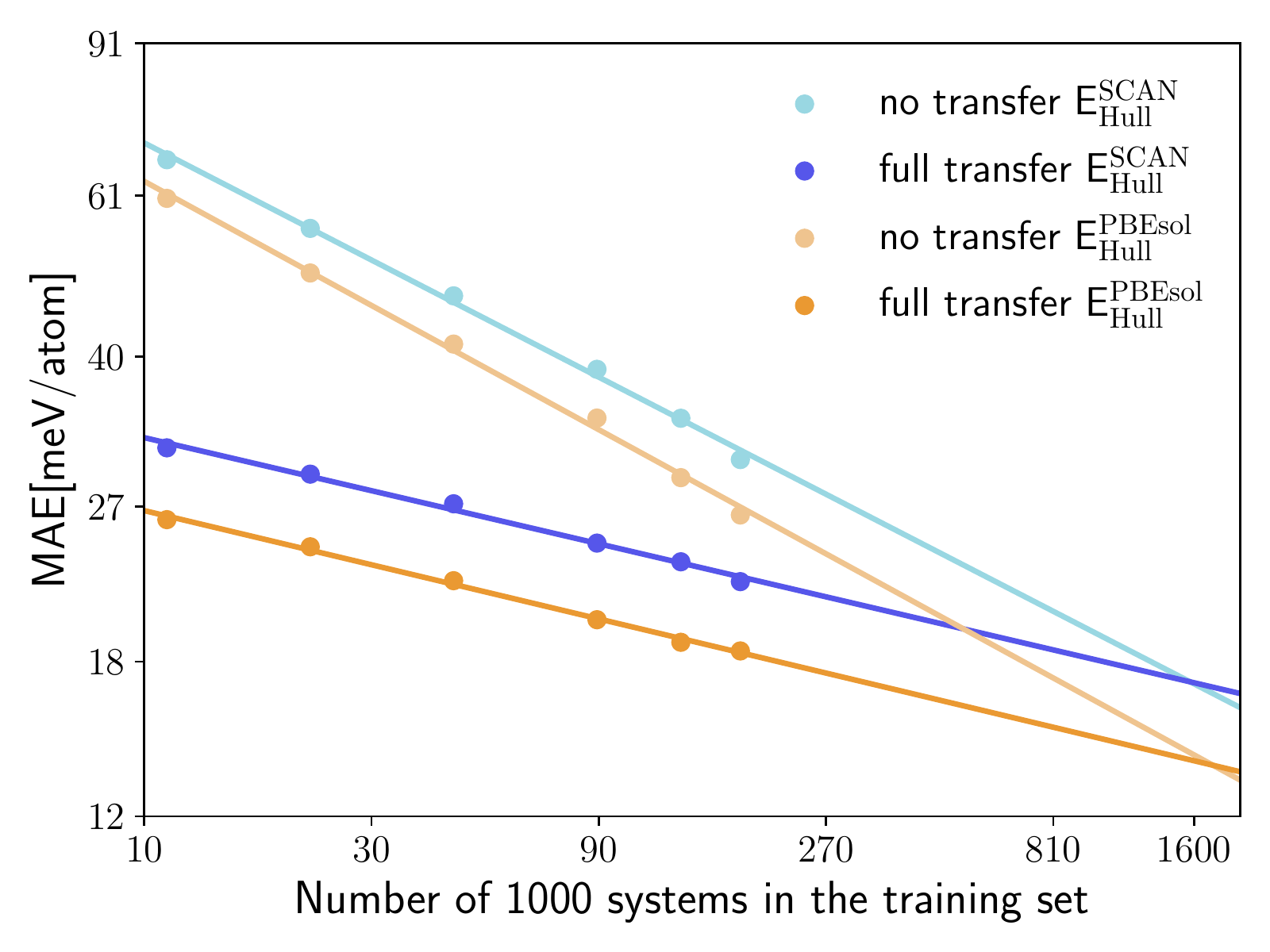}
    \caption{Log-log plot of the test MAE for the prediction of  E$_{\text{Form}}^{\text{SCAN/PBEsol}}$ (left) and E$_{\text{Hull}}^{\text{SCAN/PBEsol}}$ (right) as a function of the training set size for a model trained solely on SCAN/PBEsol data (no transfer) vs a model pre-trained on PBE data (full transfer). The lines show a log-log linear fit to the data.}
    \label{fig:errorvstrainingsetsize_ehull}
\end{figure*}

Now that we are convinced of the benefits of pre-training on a larger lower-quality dataset to speed up successive training on a higher-quality dataset, we can further inspect how the performance of transfer learning depends on the size of the training set. The log-log plots of \cref{fig:errorvstrainingsetsize_ehull} offer a clear insight into the quantity of high-quality data required in case of transfer from the pre-trained PBE model (with training of the full network on the new data) and no transfer. In the left panel of \cref{fig:errorvstrainingsetsize_ehull} we show the MAE for the prediction of SCAN and PBEsol energy distances to the convex hull E$_\text{hull}^{\text{SCAN/PBEsol}}$, while in the right panel of \cref{fig:errorvstrainingsetsize_ehull} the MAE for the prediction of SCAN and PBEsol formation energies E$_\text{Form}^{\text{SCAN/PBEsol}}$ is displayed. All models are again evaluated on the same test set. In both panels of \cref{fig:errorvstrainingsetsize_ehull} we observe that we reach chemical precision (i.e., an error below 43~meV/atom) already at 11k training systems. We can therefore conclude that fitting neural networks to computationally expensive calculations based on hybrid or double-hybrid functionals will be enabled in the near future by transfer learning. We can easily fit the points in the log-log plots with lines and extrapolate the number of training datapoints needed to achieve the same MAE with and without transfer learning. The resulting numbers are very consistent: 1.6M (E$_\text{Hull}^{\text{SCAN}}$), 1.7M (E$_\text{Hull}^{\text{PBEsol}}$), 1.5M (E$_\text{Form}^{\text{SCAN}}$) and 1.6M (E$_\text{Form}^{\text{PBEsol}}$). Similarly, we can extrapolate for how many training systems the neural network, trained without transfer learning, would have the same MAE of the best network that we have trained with transfer learning. We obtain in this case these number of systems: 613k (SCAN) and 637k (PBEsol) for predicting E$_\text{Hull}$ and 513k (SCAN) and 568k (PBEsol) for predicting E$_\text{Form}$. In other words, it is necessary at least to double the size of the training datasets of PBEsol and SCAN calculations to achieve the performance we already have with the available data.

We can now ask a similar question for the size of the pre-training data set. In fact, it can be even more interesting to assess if increasing the quantity of PBE data used for the pre-training can also improve the final MAE of the network to predict PBEsol and SCAN properties, without adding new calculations to these higher-quality datasets.
To quantify this effect we trained four neural networks, using datasets of PBE calculations with different size, to predict E$_\text{Hull}^{\text{PBE}}$. We then used these models as starting points for transfer learning to predict PBEsol and SCAN energies, following the same procedure as before. 

\begin{figure*}
    \centering
    \includegraphics[width=.45\linewidth]{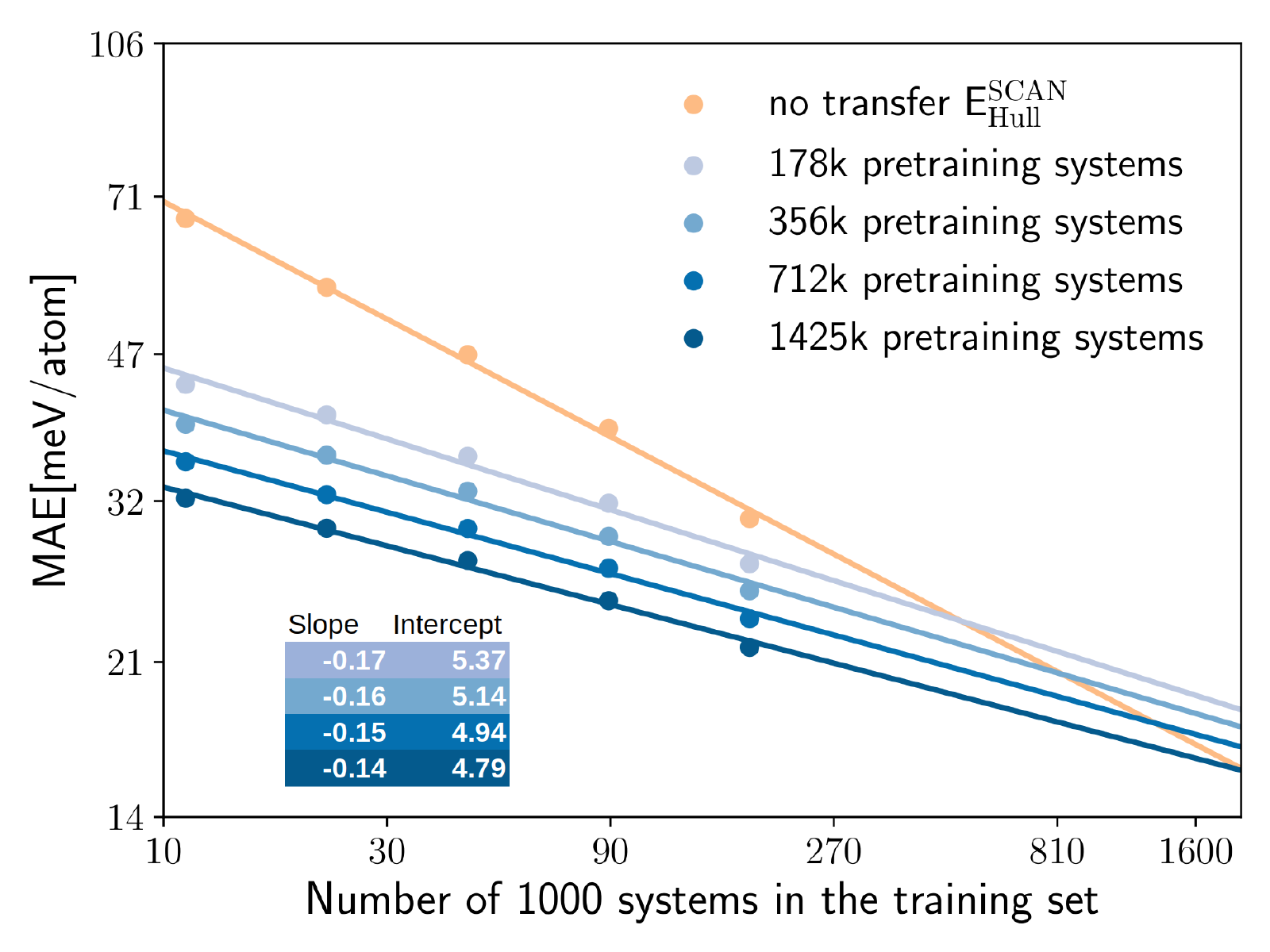}
    \includegraphics[width=.45\linewidth]{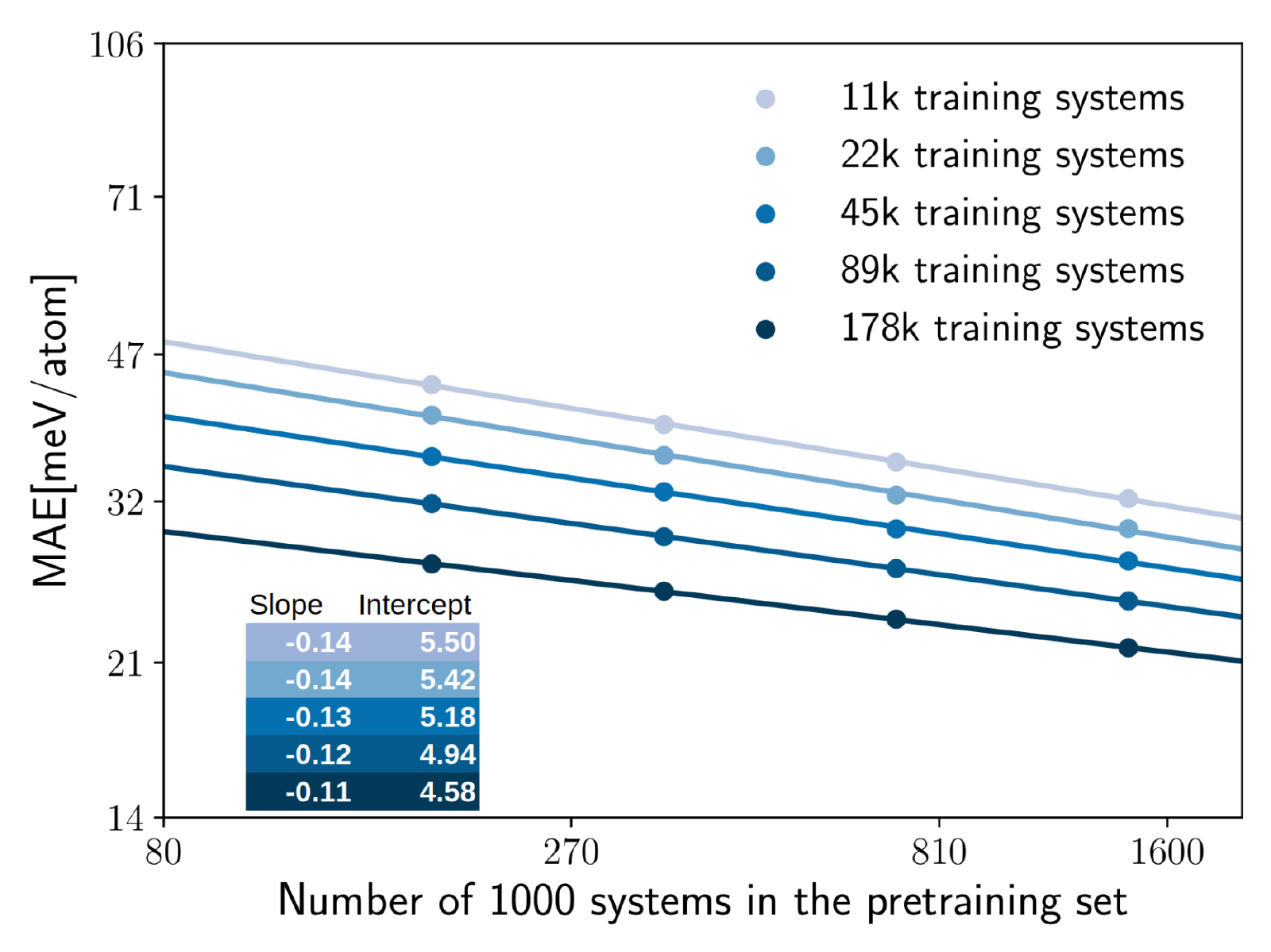}
    \caption{The left panel shows the log-log plot for the MAE on the test set for the prediction of E$_\text{Hull}^{\text{SCAN}}$~ as a function of the training set size. We consider the cases of full transfer with pretraining datasets of different size (different shades of blue) or no transfer (orange). The right panel shows the same MAE as a function of the pre-training dataset, for different values of the training datasets (different shades of blue). In the insets we indicate the slopes and the y-intercepts according to the linear fit}
    \label{fig:errorvspretrainingsetsize}
\end{figure*}

In the left panel of \cref{fig:errorvspretrainingsetsize} we can see the MAE on the test set for predicting E$_\text{Hull}^{\text{SCAN}}$, plotted on a log-log scale as a function of the number of systems in the training set. The different curves correspond to pretraining using PBE datasets of different size, while the orange curve shows the performance of the network trained without transfer learning. 

We can observe that the points draw lines with very similar slopes for pretraining datasets of different sizes. In fact, the slope decreases only slightly with the increasing size of the pre-training dataset. This behavior is expected, as when more PBE data is already given to the model, less new information can be found in the additional SCAN data. The MAE decreases significantly and consistently with the increasing size of the pretraining set. The neural network trained only on the SCAN dataset (no transfer) has the largest MAE in all cases, even when the pre-training dataset is reduced by a factor of 10.

In the right panel of \cref{fig:errorvspretrainingsetsize} we demonstrate that also the error as a function of the size of the pre-training dataset can be fitted by a line in the log-log graph. Here the slopes are smaller than in the left panel. Consequently, similar reductions of the MAE can be achieved by transfer learning at the cost of using significantly more pretraining data than training data. On the other hand, extra pre-training data can be generated with lower-level approximations at a reduced computational cost, and the pre-trained model can be used as a starting point for the training of several new models. 

It is important to comment on the cost of calculations of the same property using different DFT functionals. For example, SCAN calculations are at least five time more expensive than PBE calculations. Calculations using hybrid functionals are  even more computationally demanding. In the latter case, the possibility to use larger pre-training datasets to reduce the MAE of a predicted property becomes even more appealing and effective. 

\begin{figure*}
  \begin{tabular}{c c}
  \centering
    \includegraphics[width=.45\linewidth]{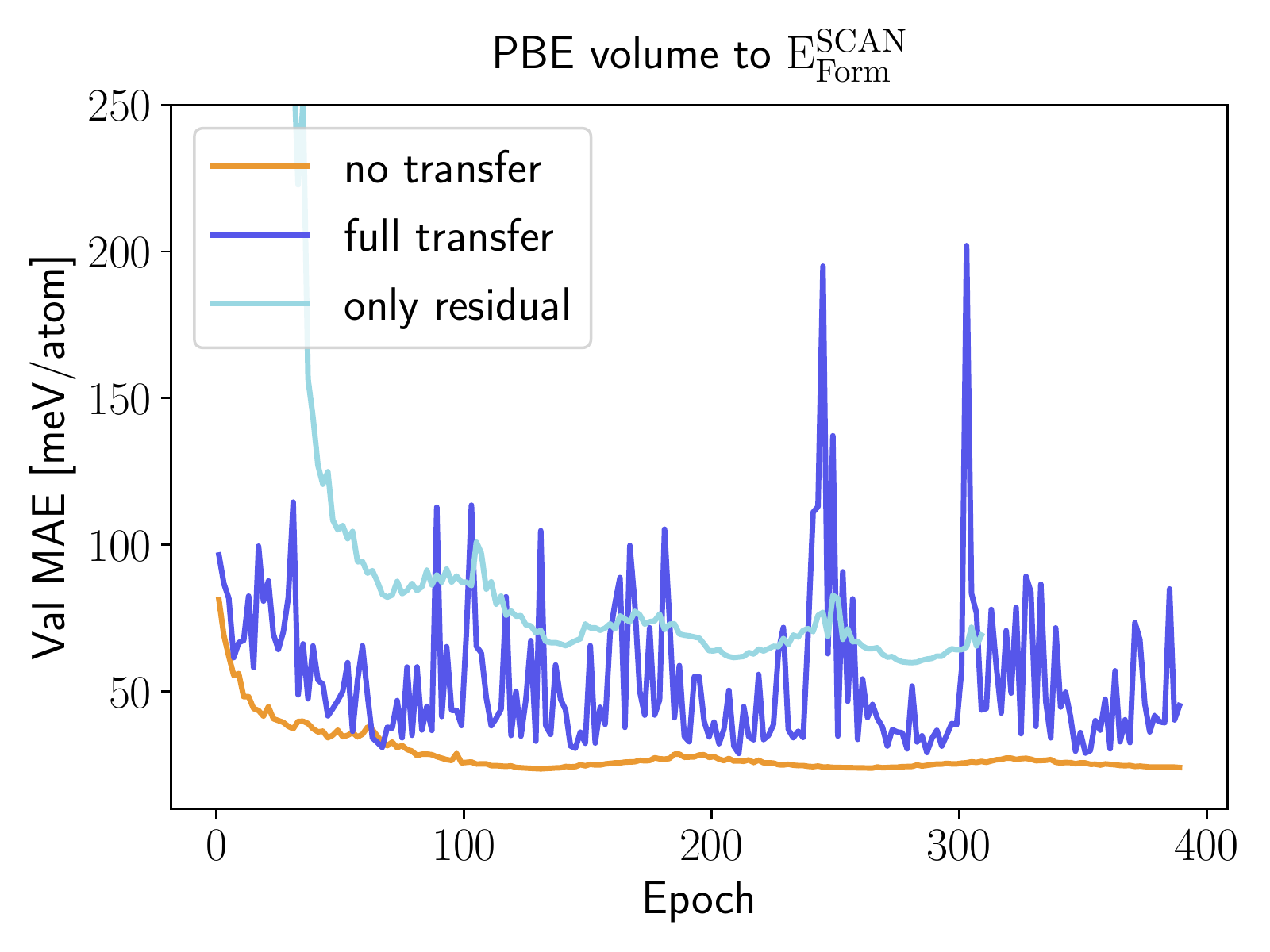}&
        \includegraphics[width=.45\linewidth]{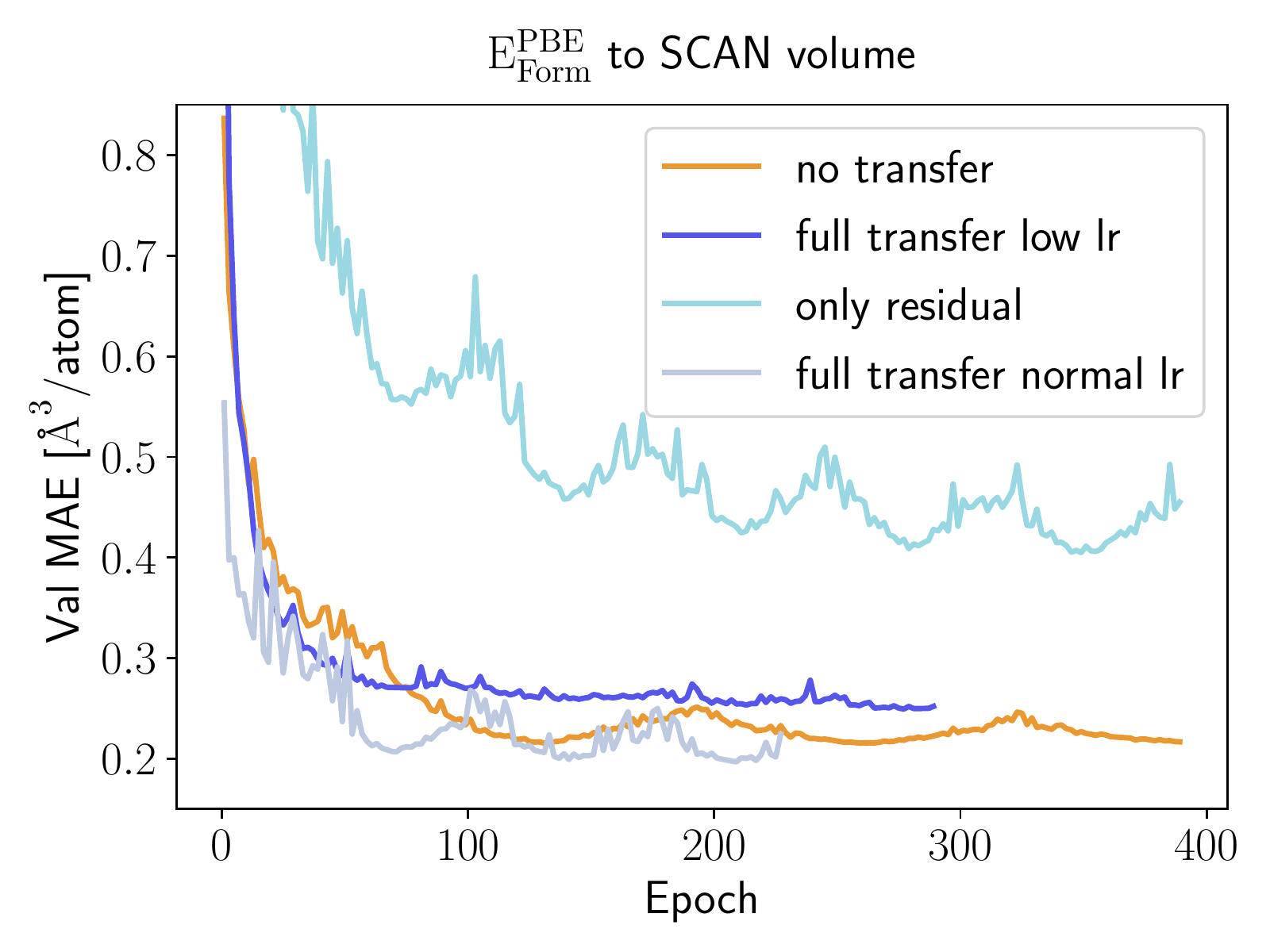}
\end{tabular}
\caption{Learning curves, i.e. MAE on the validation set as a function of the training epoch, for the prediction of PBEsol volumes with transfer learning from a model pre-trained for E$_\text{Form}^{\text{PBE}}$ (left panel) and for the prediction of E$_\text{Form}^{\text{SCAN}}$ from a model pre-trained for PBE volumes (right panel). The different curves are obtained with the training procedures described in the text: without pre-training (no transfer), with pre-training using 1.8M PBE calculations and freezing part of the weights (only residual) and with pre-training using 1.8M PBE calculations and successive full reoptimization of all weights (full transfer). In the right panel we consider full transfer with two different learning rates (lr).
}
    \label{fig:interprop-learningcurve}
\end{figure*}

Seeing the promising results of transfer learning on two datasets of calculations of the same property, we also attempted to apply transfer learning to predict different properties. We consider therefore transfer learning to predict SCAN formation energies and PBEsol volumes, starting from neural networks trained to output PBE volumes and PBE formation energies, respectively.

As we can see in \cref{fig:interprop-learningcurve} only retraining the residual network does not improve the performance of the model. This is true for both examples selected here and we can expect this to be a general rule. In fact, the strong dependence of the graph embeddings on the low-dimensional features that are good descriptors for a specific property makes this part of the network strongly property dependent. 

We observe that reoptimizing all weights of the neural network, starting from the model pre-trained to predict another PBE property, leads to a very unstable learning curve and does not produce better results than training from scratch. To enforce convergence we retrained the full network with a learning rate 10 times smaller than before. This improves marginally the situation for transfer learning for the prediction of PBEsol volumes from PBE formation energies. However, the lower learning rate leads the neural network to settle down very quickly in a suboptimal local minimum from which it is unable to escape, yielding a model that is still worse than the one obtained from training on the PBEsol dataset only. 

We have to conclude that two properties such as formation energy and optimized volume, are too far dissimilar to perform successful transfer learning. Of course, it is always possible that an extensive hyperparameter variation would improve this result. However, in this case, a hyperparameter search with the same resources should be performed also for the original model to make a valid comparison.

\begin{figure}
    \includegraphics[width=.95\linewidth]{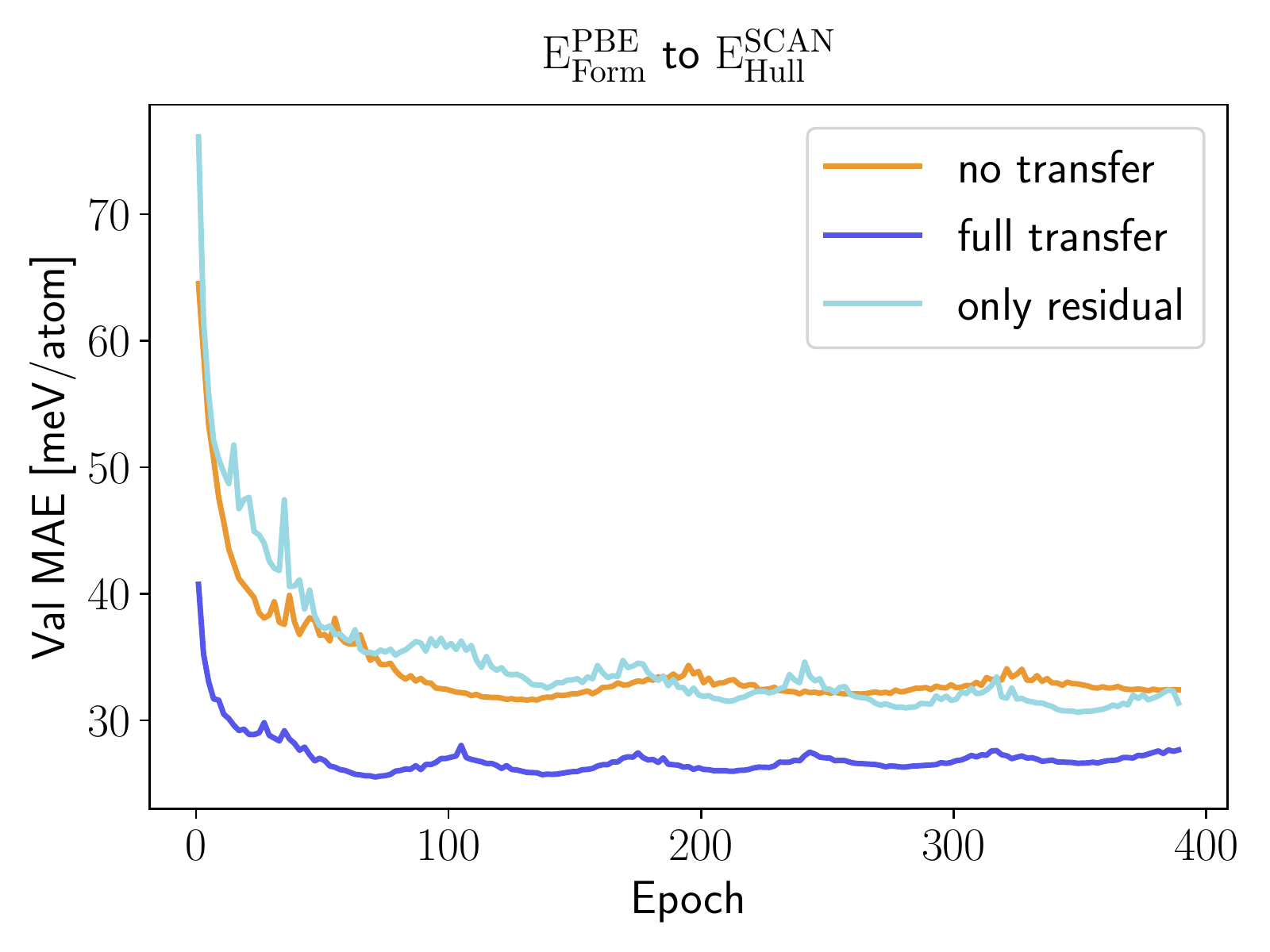}
    \caption{Learning curves, i.e. MAE on the validation set as a function of the training epoch, for the prediction of E$_\text{Form}^{\text{SCAN}}$\  with transfer learning from a neural network that predicts accurately E$_\text{Hull}^{\text{PBE}}$\. The different curves are obtained with the training procedures described in the text: without pre-training (no transfer), with pre-training using 1.8M PBE calculations and freezing part of the weights (only residual) and with pre-training using 1.8M PBE calculations and successive full reoptimization of all weights (full transfer).}
    \label{fig:pbe-e-form-to-scan-e-hull}
\end{figure}
\begin{figure}
    \centering
    \includegraphics[width=.95\linewidth]{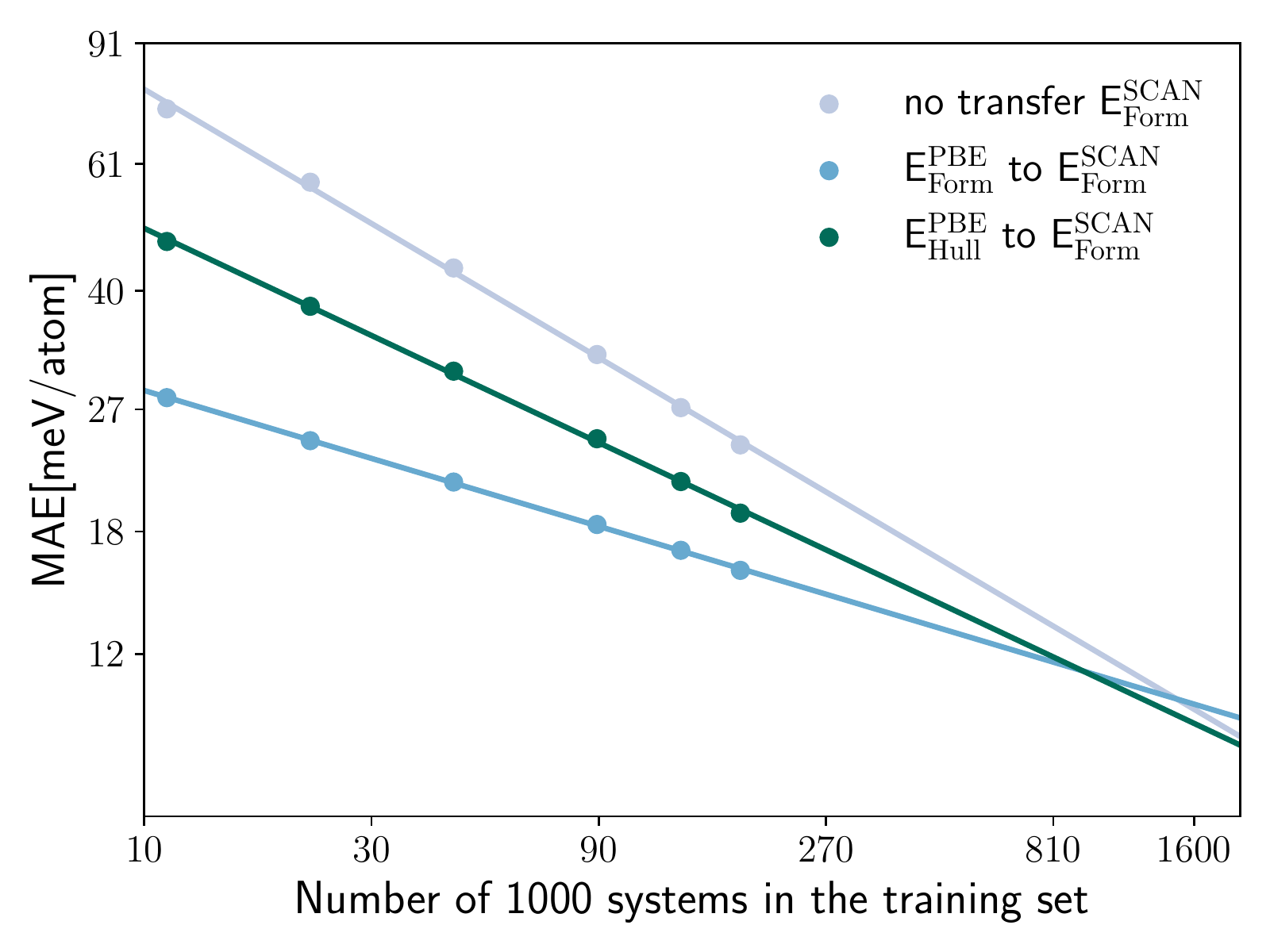}
    \caption{Log-log plot of the MAE on the validation set for the prediction of E$_{\text{Form}}^{\text{SCAN}}$~ as a function of the training set size for a model trained solely on SCAN data (no transfer), a model pre-trained on a dataset of E$_\text{Hull}^{\text{PBE}}$ calculations and a model pre-trained on E$_{\text{Form}}^{\text{PBE}}$ calculations. The lines indicate a log-log linear fit to the data.}
    \label{fig:errorvstrainingsetsize_inter}
\end{figure}

Transfer learning between a neural network that predicts PBE formation energies and a neural network that outputs SCAN energy distances to the convex hull performs better as these two properties are closely related.
In \cref{fig:pbe-e-form-to-scan-e-hull} we can see that only retraining the residual network does not bring a significant error reduction. On the other hand, the full retraining is now able to improve considerably the model performance with an error reduction of \SI{18}{\percent}. 
In \cref{fig:errorvstrainingsetsize_inter}~ we compare the MAE on the validation set for the prediction of E$_{\text{Form}}^{\text{SCAN}}$ as a function of the training set size, considering the case of no transfer learning, and two approaches for transfer learning, either starting from a neural network pre-trained on the same property calculated with PBE or on E$_{\text{Hull}}^{\text{PBE}}$. As expected, the inter-property transfer learning performs better, providing an 56--59\% larger improvement 
than the intra-property transfer learning for the whole range of considered training set sizes.

\section{Conclusions}

We demonstrated that performing transfer learning using a crystal graph neural network trained on a large (>$10^6$) dataset of less accurate but faster calculations enables efficient training of the same neural network on a smaller ($\approx 10^4-10^5$) dataset of more accurate calculations. The final prediction error is significantly lower compared to the error that would be obtained if the neural network was trained from scratch only on the smaller dataset. We demonstrated that the obtained performance improvement is consistent when we perform transfer learning for different functionals and similar electronic properties, e.g., E$_{\text{form}}$, E$_{\text{hull}}$. Thanks to transfer learning, we can assume that a training set of about $10^4$ high-quality \textit{ab initio} calculations is sufficient to obtain predictions of electronic quantities with chemical precision. High-throughput studies involving tens of thousands of calculations using advanced electronic structure methods are foreseeable in the near future. Transfer learning may therefore have a strong impact on future machine learning studies in solid-state physics, allowing prediction of properties with unprecedented accuracy.

We also demonstrated that the learning error after transfer learning decreases with increasing size of the pre-training dataset with a linear scaling in a log-log graph. Using this property, we can determine the relative size of the pre-training and training datasets, which allows us to minimize the prediction error and, at the same time, the total computational cost, given the different computational resources required for less accurate and more accurate \textit{ab initio} calculations. This is of particular interest for training universal force fields, an emerging application of machine learning that has attracted increasing attention in recent years (see, e.g., ~\cite{chen2022universal}). In fact, universal force fields are trained on PBE data so far. Since these force fields also use universal message-passing networks, we expect that transfer learning can also be easily applied to efficiently train the existing force fields to the quality of higher-fidelity functionals.  

Unfortunately, our results show that transfer learning is only effective for physically similar electronic properties. If the pre-trained property is too dissimilar, the pre-training may actually paralyze the neural network in predicting the new property. It is in that case more convenient to produce a large database of lower-quality calculations of the desired final property than to perform inter-property transfer learning.

\section{Methods}
\subsection{Data}
Our main PBE dataset consists of calculations from the Materials Project database ~\cite{jain2013commentary}, AFLOW ~\cite{curtarolo2012aflow} and our own calculations. 
This set of around two million compounds was accumulated in Ref. ~\cite{CGAT, CGATHT}.
In Ref.~\cite{dataset} we  reoptimized the geometries of 175k materials using PBEsol followed by a final energy evaluation with the PBEsol and SCAN functional as described in ~\cite{dataset}. By now we extended these datasets by another 50k randomly selected materials arriving at 225k entries.

\begin{figure*}[htb]
\centering
   \includegraphics[width=.7\linewidth]{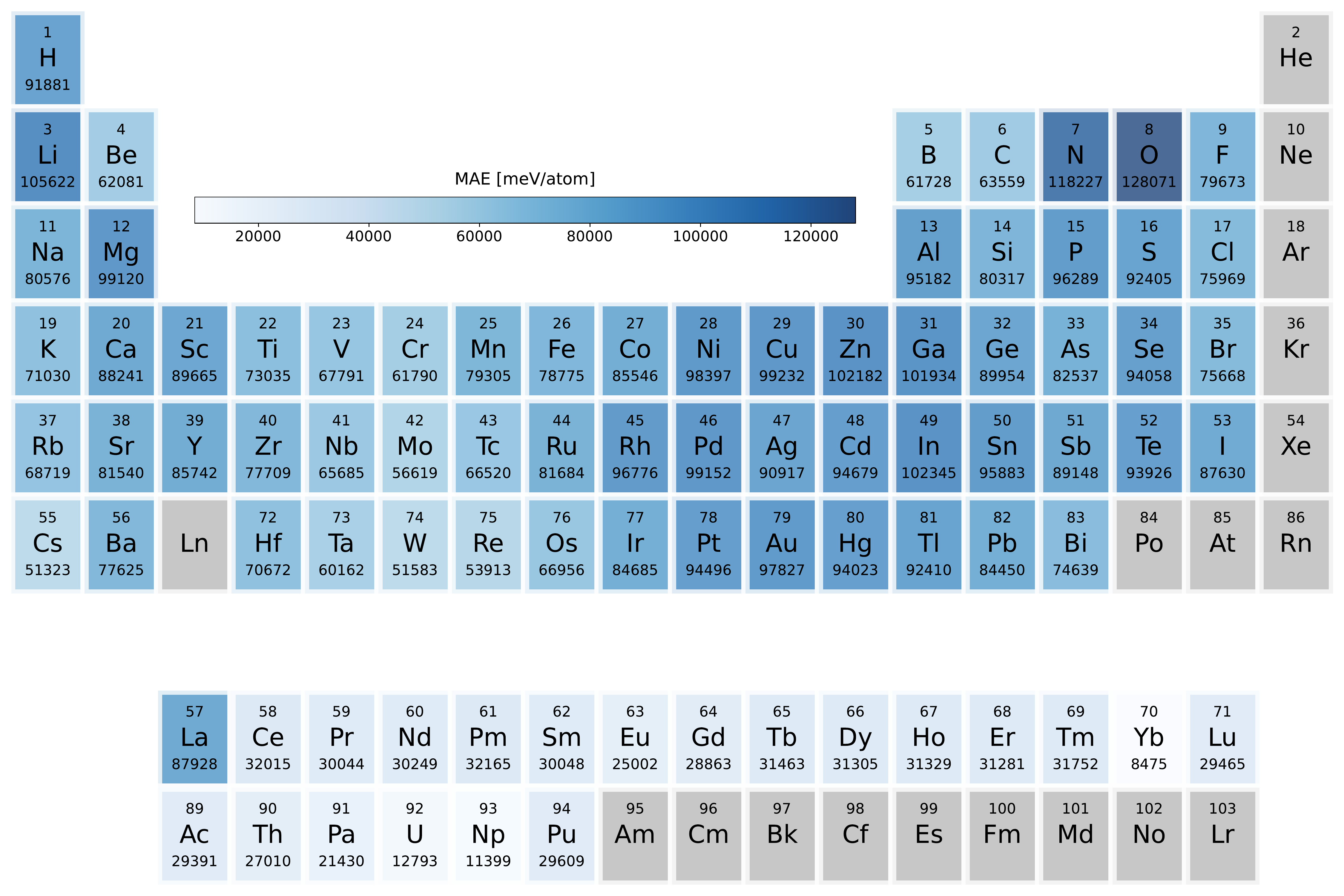}\\
(a)\\
    \includegraphics[width=.7\linewidth]{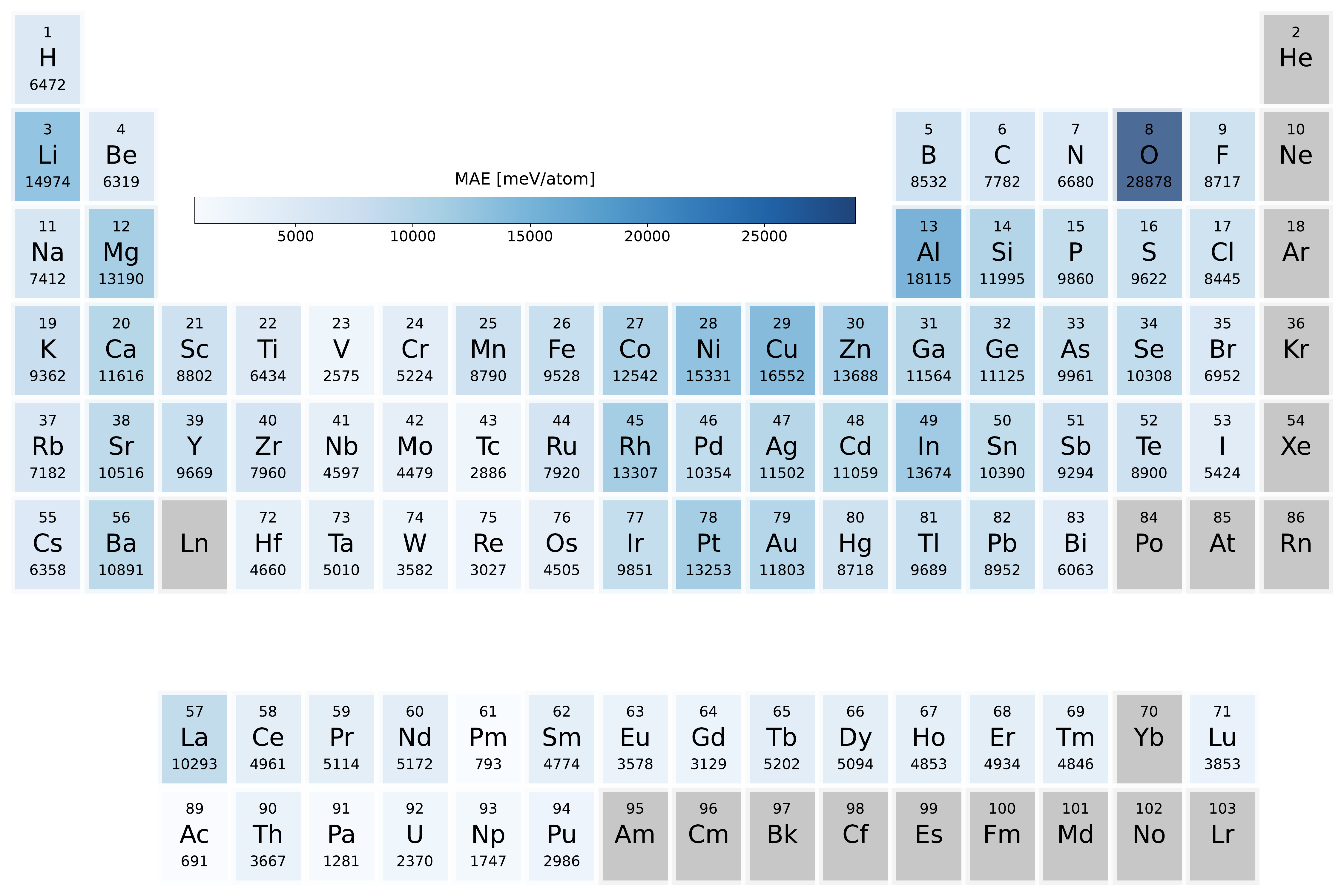}\\
    (b)
        \caption{Number of the compounds containing each elements (a) in the PBE dataset and (b) in the PBEsol/SCAN dataset.}
            \label{fig:elements}
\end{figure*}

In \cref{fig:elements}a we can see the element distribution of the PBE dataset. 
This set features a large variety of elements with oxygen and nitrogen being the most prominent followed by lithium. 
Not included are the noble gases and heavy radioactive chemical elements.
In the elemental distribution of the PBEsol and SCAN dataset, depicted in \cref{fig:elements}b, oxygen is even more prevalent while nitrogen and hydrogen appear less often. This is a result of the many stable oxides in the PBE dataset.

\begin{figure*}[htb]
  \begin{tabular}{c c}
  \centering
  \includegraphics[height=5.5cm]{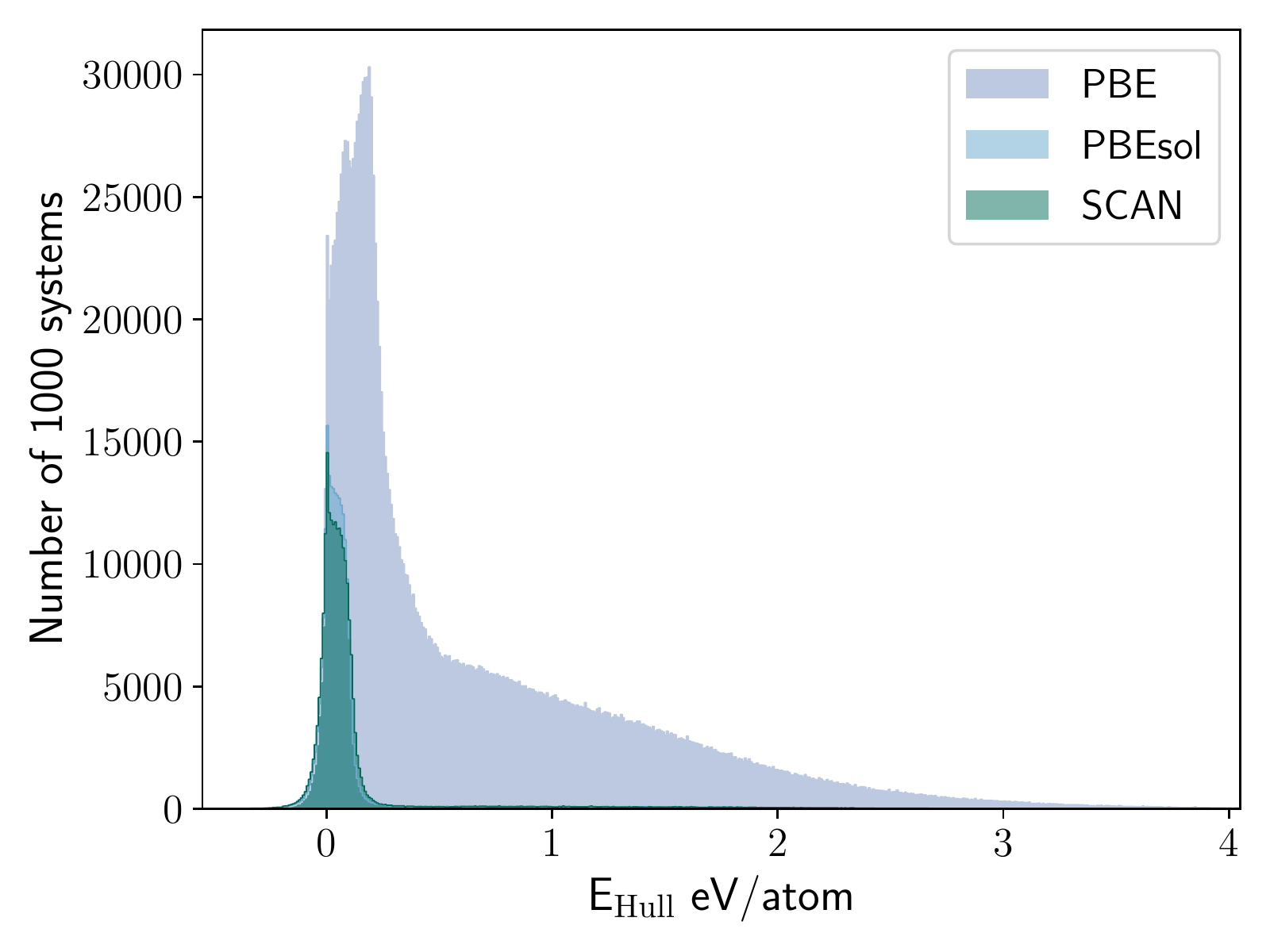}
  & \includegraphics[height=5.5cm]{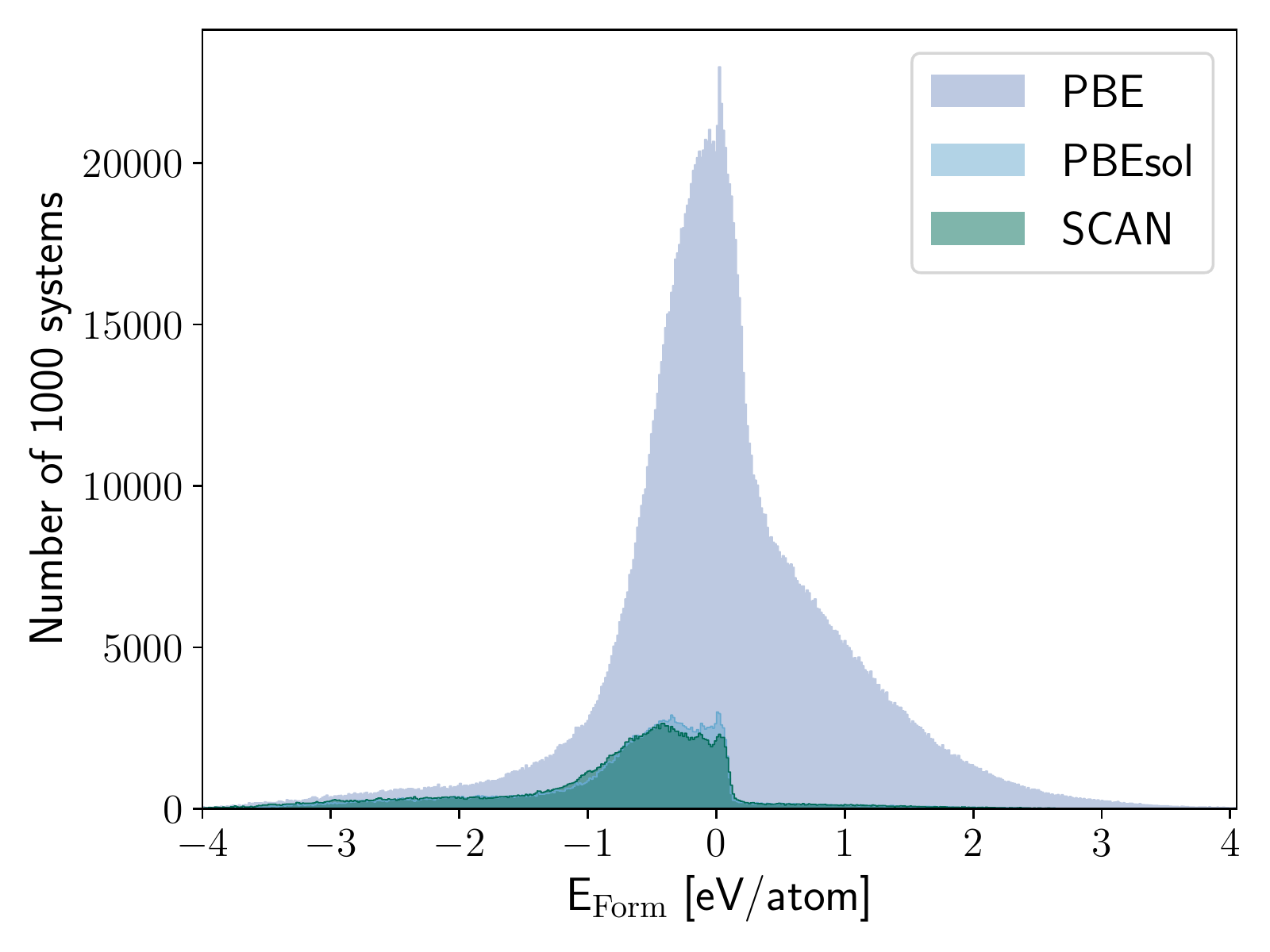} \\
    (a)   & (b) \\
  \includegraphics[height=5.5cm]{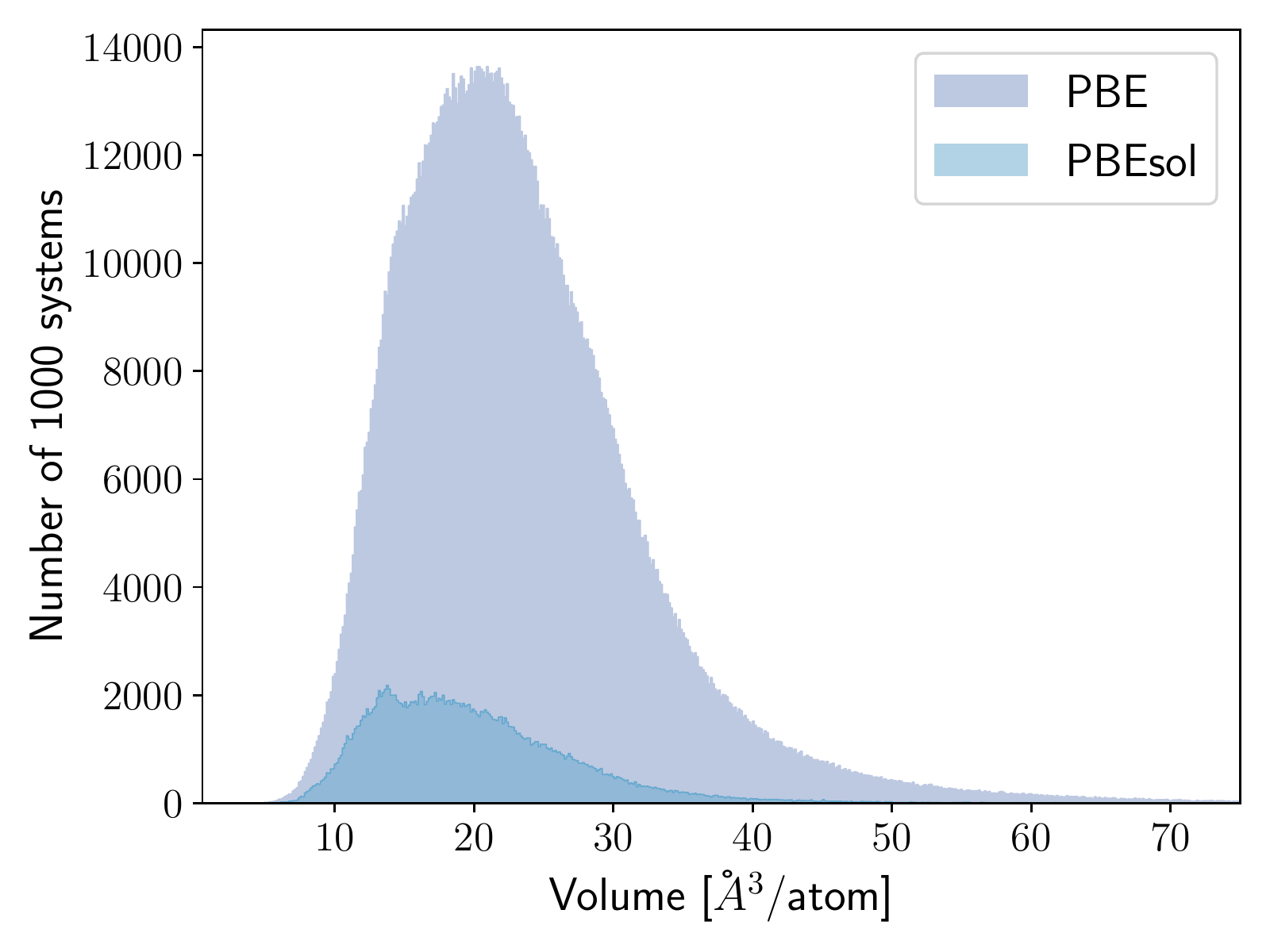} & \includegraphics[height=5.5cm]{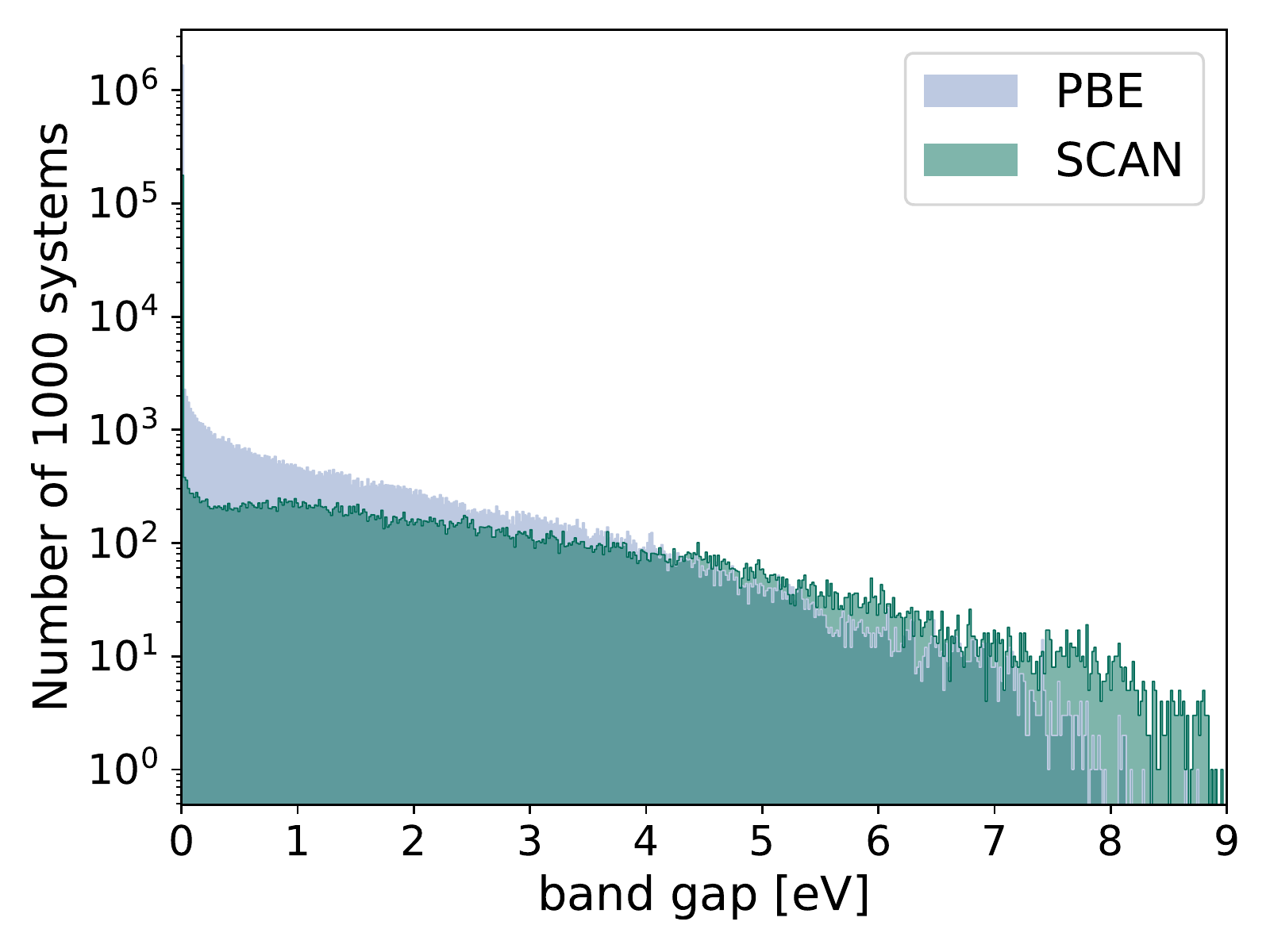} \\
     (c)  &  (d) \\
  \end{tabular}
     \caption{Histogram of (a) the distances to the convex hull and (b) the formation energies of the PBE, PBEsol and SCAN datasets. (c) Histogram of the volumes of the PBE and PBEsol dataset. (d) Histogram of the band gaps of the PBE and SCAN datasets.}
    \label{fig:hist}
\end{figure*}

In \cref{fig:hist}  we plot the distribution of the various datasets we used in this work. As expected the materials of the PBEsol/SCAN dataset are on average far more stable with only a small long tail in the E$_{\text{Form}}$\ and E$_{\text{Hull}}$\ distributions introduced by the 50k random systems. The differences in the distributions also show that the results are valid for transfer learning between datasets with rather different distributions.
The volume dataset on the other hand is very similarly distributed for PBE and PBEsol with respective medians/means/standard deviations of 22.1/23.5/9.2 \AA$^3$/atom and 18.7/19.9/7.4 \AA$^3$/atom. The slightly higher median and mean of the PBE dataset are expected due to the underbinding of the PBE that is somewhat corrected by PBEsol.
The distribution of the band gaps was already discussed earlier but the main difference is the percentage of metals that is roughly 4 times larger in the PBE dataset.

\subsection{Crystal graph attention networks}

Crystal graph attention networks were developed in Ref.~\cite{CGAT} for the discovery of new stable materials. 
Using the periodic graph representation of the crystalstructure, the networks apply an attention based message passing mechanism. By using solely the graph distance of the atoms to their neighbors as edge information, CGATs can perform precise predictions based on unrelaxed geometries.  Follow the notation of Ref.~\cite{CGAT}, we label the embedding of the $i_{\text{th}}$\ node, i.e. atom, at time steps t of the message passing process as $h_i^t$\ and the respective edge embedding to the atoms j as $e^t_{ij}$. $\text{FCNNN}^{t,n}_a$\ is the network of the $n_{\text{th}}$\ attention head at message passing step $t$\ and $\text{HFCNN}^t_{\theta^t_g}$\ a hypernetwork depending on the difference between the embedding at step t and the previous step.
Using this notation we arrive at the following equations for the updates of the node embeddings:
\begin{subequations}
\begin{gather}
    \mathbf{\mathbf{s}}^{t,n}_{ij}= \text{FCNNN}^{t,n}_a(\mathbf{h}^t_i||\mathbf{h}^t_j||\mathbf{e}_{ij})\\
    \mathbf{a}^{t,n}_{ij} =\frac{\exp(s^{t,n}_{ij})}{\sum_j \exp(s^{t,n}_{ij})} \\
    \mathbf{m}^{t,n}_{ij}=\text{FCNN}^{t,n}_m(\mathbf{h}^t_i||\mathbf{h}^t_j||\mathbf{e}_{ij}).\\
\label{eq:combine}
    \mathbf{h}^{t+1}_i = \mathbf{h}^{t}_i +\text{HFCNN}^t_{\theta^t_g}\left(\frac{1}{N}\sum_n\sum_j\mathbf{a}^n_{ij} \mathbf{m}^n_{ij}\right). 
\end{gather}
\end{subequations}

The edges are updated similarly:
\begin{subequations}
\begin{gather}
\mathbf{s}^{e,n}_{ij}= \text{FCNNN}^n_a(\mathbf{h}^t_i||\mathbf{h}^t_j||\mathbf{e}^t_{ij})\\
\mathbf{a}^{e,n}_{ij} =\frac{\exp(\mathbf{s}_{ij})}{\sum_n \exp(\textbf{s}^n_{i})}\\
\mathbf{m}^{e,n}_{ij}= \text{FCNNN}^n(\mathbf{h}^t_i||\mathbf{h}^t_j||\mathbf{e}^t_{ij})\\
\mathbf{e}^{t+1}_{ij} = \mathbf{e}^{t}_{ij}+ \text{FCNN}^{n,t}_{\theta^t_g} \left( \underset{n}{||}\mathbf{a}^{e,n}_{ij} \mathbf{m}^{e,n}_{ij}. \right)
\end{gather}
\end{subequations}
In parallel a ROOST ~\cite{goodall2019predicting} model calculates a representation vector of the composition that is used as a global context vector and is concatenated with the final node embeddings. Lastly, an attention layer calculates the embedding for the whole crystal structure. Then a residual network transforms the graph embedding into the prediction. We used the following hyperparameters:
 AdamW;
  learning rate:  0.000125 (0.0000125 for transfer learning);
  starting embedding: matscholar-embedding\cite{matscholar};
  nbr-embedding-size: 512;
  msg-heads: 6;
  batch-size: 512;
  max-nbr: 24;
  epochs: 390;
  loss: L1-loss;
  momentum: 0.9;
  weight-decay: 1e-06;
  atom-fea-len: 128;
  message passing steps: 5;
  roost message passing steps: 3;
  other roost parameters: default;
  vector-attention: True;
  edges: updated;
  learning rate: cyclical;
  learning rate schedule: (0.1, 0.05);
  learning rate period: 130 (70 for transfer learning, 90 for pre-training during the experiments shown in \cref{fig:errorvspretrainingsetsize});
  hypernetwork: 3 hidden layers, size 128;
  hypernetwork activation function: $\tanh$;
  FCNN: 1 hidden layer, size 512;
  FCNN activation function: leaky RELU~\cite{LIEW2016718};
  Nvidia apex mixed precision level O2 (almost FP16).
\section{Data availability statement}
This study was carried out using publicly available data from \url{https://doi.org/10.24435/materialscloud:ka-br} and \url{https://doi.org/10.24435/materialscloud:m7-50}. Additional SCAN and PBEsol data will be added to \url{https://doi.org/10.24435/materialscloud:ka-br}.

\section{Code availability statement}
The CGAT code is available on  \url{https://github.com/hyllios/CGAT}. The trained models resulting from all the experiments are available upon request.

\section{Author  Contributions}
JS and NH performed the training of the machines and the machine learning predictions of the distance to the hull; MALM performed the DFT high-throughput calculations; JS, SB and MALM  directed the research; all authors contributed to the analysis of the results and to the writing of the manuscript.

\section{Competing  Interests}
The authors declare that they have no competing interests.

\section{Acknowledgements}
The authors gratefully acknowledge the Gauss Centre for Supercomputing e.V.
(www.gauss-centre.eu) for funding this project by providing computing time on
the GCS Supercomputer SUPERMUC-NG at Leibniz Supercomputing Centre
(www.lrz.de) under the project pn25co. This project received funding from the European Union’s HORIZON-MSCA-2021-DN-01 program under grant agreement number 101073486 (EUSpecLab).

\section{References}
\bibliography{ref.bib}

\end{document}